\documentclass[aps,prd,showpacs,nofootinbib]{revtex4}
\usepackage{graphicx}
\begin{document}
\baselineskip=20pt
\title{Mixed high energy neutrinos from cosmos}
\author{H. Athar}
\affiliation{Physics Division, National Center for Theoretical
         Sciences, 101 Section 2, Kuang Fu Road,
         Hsinchu 300, Taiwan,
         E-mail: athar@phys.cts.nthu.edu.tw}
\date{\today}
\begin{abstract}

Production of the expected high energy neutrino flux with energy 
greater than tens of thousands of GeV in some
 astrophysical sites such as the galactic plane as well as
the centers of some distant galaxies is reviewed. The expected changes in these
neutrino fluxes because of neutrino oscillations during their
propagation to us are described. Observational signatures for these
neutrino fluxes with and without neutrino oscillations are
discussed.

\end{abstract}
\pacs{13.85.Tp, 95.85.Ry, 14.60.Pq, 95.55.Vj}
\maketitle
\section{General Introduction}
    In the standard model of the electro weak interactions, the
lepton masses and the values of other parameters such as weak
mixing angle, couplings, etc. are arbitrary and are therefore
determined by experiments.
These parameters are independent of each other and can not be
determined uniquely, while the neutrino is taken to be massless
because of maximal parity violation. The masslessness of the
neutrino however does not follow from any other theoretical ground unlike
the local gauge invariance for the photon \cite{Fukugita:wx,Kim:dy}.

    In order to search for physics {\tt beyond} the standard model of particle 
 physics, light
neutrino masses are incorporated in the extensions of the
standard model. 
 Empirically, one places finite non-vanishing upper bounds
 on measurable neutrino masses. Presently, there is some indirect
    empirical evidence for the masslessness of the 
neutrino \cite{Pakvasa:2003zv}.

     Massive neutrinos quite likely mix.
    The mixing of quarks is an established fact and because of
    quark-lepton symmetry, it is natural to assume that leptons
    exhibit mixing as well. An additional argument, in this sense,
    is provided by grand unification models, in which quarks and
    leptons are described in a unified manner.

    Mixing means that $\nu_{e}$, $\nu_{\mu}$ and $\nu_{\tau}$,
i.e., the states created in weak interactions, are different from
the states $\nu_{1}$, $\nu_{2}$ and $\nu_{3}$ that have definite
masses. The neutrinos $\nu_{e}$, $\nu_{\mu}$ and $\nu_{\tau}$ are
orthogonal combinations of $\nu_{1}$, $\nu_{2}$ and $\nu_{3}$ with
different phases between them. In addition, the sterile neutrinos
may mix with these neutrinos.

    Neutrino mixing has, as its consequence, {\tt neutrino
    oscillations}, i.e., the process of periodic (complete or
    partial) conversion of neutrinos of one type into another, for
    instance, $\nu_{e}\to \nu_{\mu}\to \nu_{e}\to ... $ . The
    components $\nu_{i}$ of a mixed neutrino have different masses
    and hence different phase velocities. It follows that the
    phase difference caused by the mass difference between the
    $\nu_{i}$ vary monotonically during the propagation. This
    phase change manifests itself as neutrino oscillations.

If neutrino oscillations do occur in vacuum, matter can enhance
their depth (probability amplitude) up to a maximal value
\cite{Mikheev:qk}. That is, a monotonic change of density may lead
to resonant conversions between various neutrino flavors. This
follows from the fact that when neutrinos propagate through a
monotonically changing density medium, $\nu_{e}$ and $\nu_{\mu}$
($\nu_{\tau}$) feel different potentials, because $\nu_{e}$
scatters off electrons via both neutral and charged currents,
whereas $\nu_{\mu}(\nu_{\tau})$ scatters off electrons only via
the neutral current. This induces a coherent effect in which
maximal conversion of $\nu_{e}$ into $\nu_{\mu}$ take place (even
for a rather small intrinsic mixing angle in the vacuum), when the
phase difference arising from the potential difference between the
two neutrinos cancel the phase caused by the mass difference in
the vacuum \cite{Wolfenstein:1977ue}.

 During nearly past half a century, the empirical search for neutrinos has
spanned roughly six orders of magnitude in neutrino energy $E$, from $\sim 10^{-3}$ GeV
up to $\sim  10^{3}$ GeV. The lower energy edge corresponds to the 
Solar neutrinos, whereas the upper energy edge corresponds to the
Atmospheric neutrinos. A detailed early description of the Solar
neutrino search can be found in \cite{Bahcall:gw}, whereas for recent status, see, for instance 
  \cite{Giunti:2003tc}. The aspects of
neutrino production in Atmosphere of earth related to neutrino oscillation
studies are recently reviewed in \cite{Gaisser:2002jj}. The intermediate energy range
corresponds to terrestrial neutrinos such as from reactors (and accelerators) and the  
 Supernova neutrinos. Thus, obviously either going in
energy range below these values or {\tt above} are the available
frontiers.   For a general introduction of the
possibility of having neutrinos with energy $> 10^{3}$
GeV, see \cite{Bahcall:iz}. The upper energy edge for these high 
 energy neutrinos is limited only by the concerned experiments. 
 More detailed general discussions in the context of high energy 
 neutrinos can be found in  \cite{Ginzburg:sk,Learned:sw}.
 Despite the availability of the
somewhat detailed discussion of progress cited in the last reference, the field of high 
energy neutrino astrophysics is still passing through its 
initial stage of development.

Main  terrestrial empirical highlights include the early discovery of the
$\bar{\nu}_{e}$, the existence of three light stable
neutrinos flavors $\nu_{e}$, $\nu_{\mu}$ and $\nu_{\tau}$ as well as the recent 
 (tentative) evidence for the $\bar{\nu}_{e} \to \bar{\nu}_{\mu}$
oscillations. The extra-terrestrial highlights include the
observation of Solar neutrinos and the $\nu_{e} \to \nu_{\mu, \tau}$
oscillations for these neutrinos, the same from Atmosphere of earth with 
 Atmospheric $\nu_{\mu}\to 
\nu_{\tau}$ oscillations. The confirmation of role of neutrinos
in dynamics of supernova core collapse via SN 1987A also needs to be
mentioned here \cite{Athar:2002uj}.

The above highlighted empirical search has thus already given us quite
useful insight into neutrino intrinsic properties such as
mass and mixing. 
Massive neutrinos and their associated properties such as Dirac or
Majorana character of their mass, their mixings (and magnetic
moments) can have important consequences in astrophysics. In this
 review, I shall discuss some of the selected  consequences and the
constraints implied by these consequences on neutrino properties
as well as the insight that one may gain about the nature of the
astrophysical (or/and cosmological) sites and the interactions that 
 produce these high energy neutrinos.
 The explanation of observed Solar $\nu_{e}$ deficit relative to its 
production value in the core of the Sun, via $\nu_{e}\to \nu_{\mu }, \nu_{\tau} $ 
conversion occurring inside the Sun is an impressive 
example in this context \cite{Smirnov:2002in}.

The general plan of this mainly pedagogical review is as follows. 
 In Section IIA and IIB,   
 the presently envisaged motivations and absolute levels of the 
 expected high energy neutrino
flux from some representative examples of 
 cosmos around us such as our galactic plane are presented. 
 This includes also the one 
 arising from the interaction of ultra
high energy cosmic ray flux ($E \geq 10^{9}$ GeV) with the matter
and radiation inside the sources (such as center of our and other
galaxies) of as well as during propagation of ultra high energy
cosmic ray flux to us.
In Section IIC, the effects of neutrino flavor mixing for high energy
neutrino flux are discussed, whereas in Sect. IID, main high energy 
neutrino interactions relevant for possible future observations are 
described along with a discussion of possibility of neutrino 
flavor identification. Section III gives the  summary and conclusions.  
\section{High energy neutrinos from cosmos}
\subsection{Introduction}

 Neutrinos with $E > 10^{3}$ GeV are expected to mainly arise from the
interaction of ultra high energy cosmic rays considered to be protons ($p$) here 
 with the matter ($p$) and/or
radiation ($\gamma $) present in cosmos. Examples of the astrophysical sites
where these interaction may occur include the galactic plane,
other sites within our galaxy as well as distant sites such as centers of
 nearby active galaxies (AGNs) 
and sites for gamma ray bursts (GRBs). Searching for high energy neutrinos thus 
can in turn also constrain the particle identity of ultra high energy cosmic rays.

 As mentioned in the general introduction, here plan  is to briefly review the
present motivations of study of these 
 high
energy neutrinos. 
 Though, so far there is no observation of neutrinos with
energy greater than few thousand GeV, whose origin can not be
associated with the Atmosphere of earth, nevertheless, somewhat
optimistically speaking, given the current status of high energy
neutrino detector developments and the absolute levels of predicted high
energy neutrino fluxes, it is expected that possibly the
first evidence of high energy neutrinos may appear within this decade.

A main motivation of high energy neutrino search 
is the quest of the microscopic understanding of the nature and origin
of observed ultra high energy cosmic rays, namely the presently open questions
 such as 
whether they are protons, photons, neutrinos, heavy nuclei such as
iron nuclei or some other particles suggested beyond the standard
model of particle physics, and where and how they are produced or
accelerated. A positive observation of high energy neutrinos can raise
the possibility of simultaneous explanation of observed high
energy photons ($E_{\gamma}\simeq 10^{3}$ GeV) and ultra high energy 
 cosmic rays as a result of
hadron acceleration and interaction in the presently expanding
universe.

 Neutrinos with energy $> 10^{3}$ GeV can act as {\tt probes}
  of the ultra high energy
phenomena observed in the Universe. Unlike photons and charged
particles such as protons and heavy nuclei, which can be absorbed or deflected by dust, 
 other
intervening matter or magnetic fields, neutrinos can more easily
reach the earth because of their weak interactions
with matter particles. It is therefore hoped that such neutrinos can
provide information about the astrophysical (or/and cosmological)  
 sources that will be
complementary to inferences based on visual observations. A better
understanding of the interactions involved in neutrino production and
a more accurate estimate of resulting neutrino fluxes could entail
important consequences. Among these are insights into 
 intrinsic properties of neutrinos  such as mass and mixing 
 \cite{Athar:2000yw}, and the possible role of gravity on
neutrino propagation in astrophysical environments \cite{Athar:2000ak}.
However, it all depends on the existence of a sizable high energy neutrino
flux.  Assuming an existence of a sizable high energy neutrino flux,
    several of the other neutrino intrinsic properties as well as the
    useful information about the source producing these neutrinos
    can be obtained, at least in principle. These include testing neutrino decay 
 hypothesis \cite{Keranen:1999nf},
    constraints on neutrino magnetic moment \cite{Mughal:hr}, 
 quantum gravity effects on
neutrino propagation \cite{Athar:1999gx}, tests of possible violation of
equivalence principle by neutrinos \cite{Minakata:1996nd}, as well as 
information on different properties of relic neutrinos
\cite{Weiler:1982qy}. Also possibly enhancement in neutrino
nucleon interaction cross section because of various new physics effects may be
constrained \cite{Tyler:2000gt}. An early attempt to constrain 
the neutrino nucleon interaction cross section is discussed in \cite{Berezinsky:kz}.

The relevant average physical picture in AGNs is as follows. 
Some galaxies ($\sim 1\%$) have quite bright centers. The photon
luminosity of these galaxies typically reach ($10^{44}-10^{48}$)
erg s$^{-1}$. These galaxies are typically several Mpc away from us 
 (where 1 pc $\simeq 3\cdot 10^{18}$ cm). In
 general, AGNs
refer to these bright and compact central regions, 
 which may extend up to several pc in the center. These central
compact regions have the remarkable property of being much more
luminous than the rest of the entire galaxy.  It is hypothesized
that the existence of a super massive black hole with mass, $m_{\rm BH}\,
\sim (10^{6}-10^{10})\,
 m_{\odot}$, where $m_{\odot}\sim 2\cdot 10^{33}$ g,  may explain the observed brightness as this
super massive black hole captures the matter around it through
accretion. This super massive black hole is presently hypothesized to be
formed by the collapse of a cluster of stars. Some AGNs give off a
jet of matter that stream out from the central compact
 region  in a
 transverse plane and produce hot spots when the jet strikes the
surrounding matter at its other ends. During and after accretion, the (Fermi)
accelerated ultra high energy protons may collide with other protons and/or with the
ambient photons in the vicinity of an AGN or/and in the
associated jets/hot spots to produce unstable hadrons (such as $\Delta $). These
unstable hadrons decay mainly into neutral and charged pions. The
 neutral pions further decay dominantly into
photons and thus may explain a large fraction of the observed
brightness, 
 whereas the charged pions
mainly decay into neutrinos. AGNs, therefore,  have been targeted
as one likely source of high energy neutrinos. Currently,
the
 photohadronically ($p\gamma $) produced diffuse
flux of high energy neutrinos originating from AGNs
dominate over the flux from other sources above the relevant
Atmospheric neutrino background typically for $E\, \geq 10^{6}$ 
 GeV \cite{Protheroe:1998dm,Halzen:1998mb}.
For further reading on astrophysical super massive black holes, see
\cite{Ferrarese:2002vg}.

    Recently, fireballs are suggested as a possible production scenario 
 for gamma ray bursts as well as high energy neutrino bursts at the site 
\cite{Waxman:1997ti}. Though, the origin of these  gamma ray
burst fireballs is not yet understood, the observations suggest
that generically a very compact source of linear scale $\sim
10^{7}$ cm through internal or/and external shock propagation
produces these gamma ray bursts (as well as burst of high energy
 neutrinos) mainly in $p\gamma $ interactions. 
 Typically, this compact source is hypothesized
to be formed possibly due to merging of binary neutron stars or
due to collapse of a super massive star. Thus, 
fireballs have also been suggested as a probable scenario for the
observed gamma ray
 bursts, and they too are expected to emit neutrinos with energies in excess
  of thousands of GeV, possibly above the Atmospheric neutrino background. For a recent review,
see \cite{Dermer:2000yd}. 

	A nearby and more certain source of high energy neutrinos is our galactic plane.
The incoming ultra high energy cosmic ray protons interact with the ionized hydrogen 
clouds there and can produce high energy neutrinos in $pp$ interactions. Present
estimates indicate that the diffuse galactic plane muon neutrino flux can
{\tt dominate} over the Atmospheric one for $E > 10^{5}$ GeV. Other sources within the 
galaxy such as galactic micro quasars may also produce high energy muon neutrino 
flux around this energy. 
\subsection{Expected neutrino production}
Several types of interactions and the resulting unstable 
particles can in principle give rise to high energy neutrino 
 flux in cosmos. For definiteness, here I shall consider only $p(\gamma,p)$
 interactions to illustrate some examples of presently 
 envisaged main source interactions and to classify the
 expected high energy neutrino production sites accordingly.

    A presently favorable astrophysical (or bottom up) scenario for high energy neutrino production
 is that the
\begin{table}
\caption{Comparison of the cross sections and average fraction of incident high energy carried by 
 neutrinos for the three high energy neutrino production 
 interactions discussed in the text
 at $\sqrt{s}\sim 1.2  \, \, {\rm GeV}$.}
\begin{tabular}{|c|c|c|c|}
\hline
\hline
\label{tableone}
 Interaction & $\sigma $(mb)& Average fraction of incident high energy carried by neutrinos\\
 \hline
  $p\gamma \to N\pi^{\pm}$ & $\leq  5\cdot 10^{-1}$ & $E_{\nu}/E_{p}\leq 5$ \%\\
 $pp\to N\pi^{\pm}$ & $\sim 3\cdot 10^{1}$& $E_{\nu}/E_{p}\leq 5$ \%\\
 $\gamma \gamma \to \mu^{+} \mu^{-}$ & $ < 10^{-3}$ & $E_{\nu}/E_{\gamma}\leq 30$  \%\\
\hline
\hline
\end{tabular}
\end{table}
    observed ultra high energy cosmic rays beyond GZK cutoff (see
    later) are dominantly protons and that the observed high
    energy photon flux  can be
    associated with these. On the other hand, an unfavorable 
    scenario is that the ultra energy cosmic rays are 
 dominantly other than protons   and that
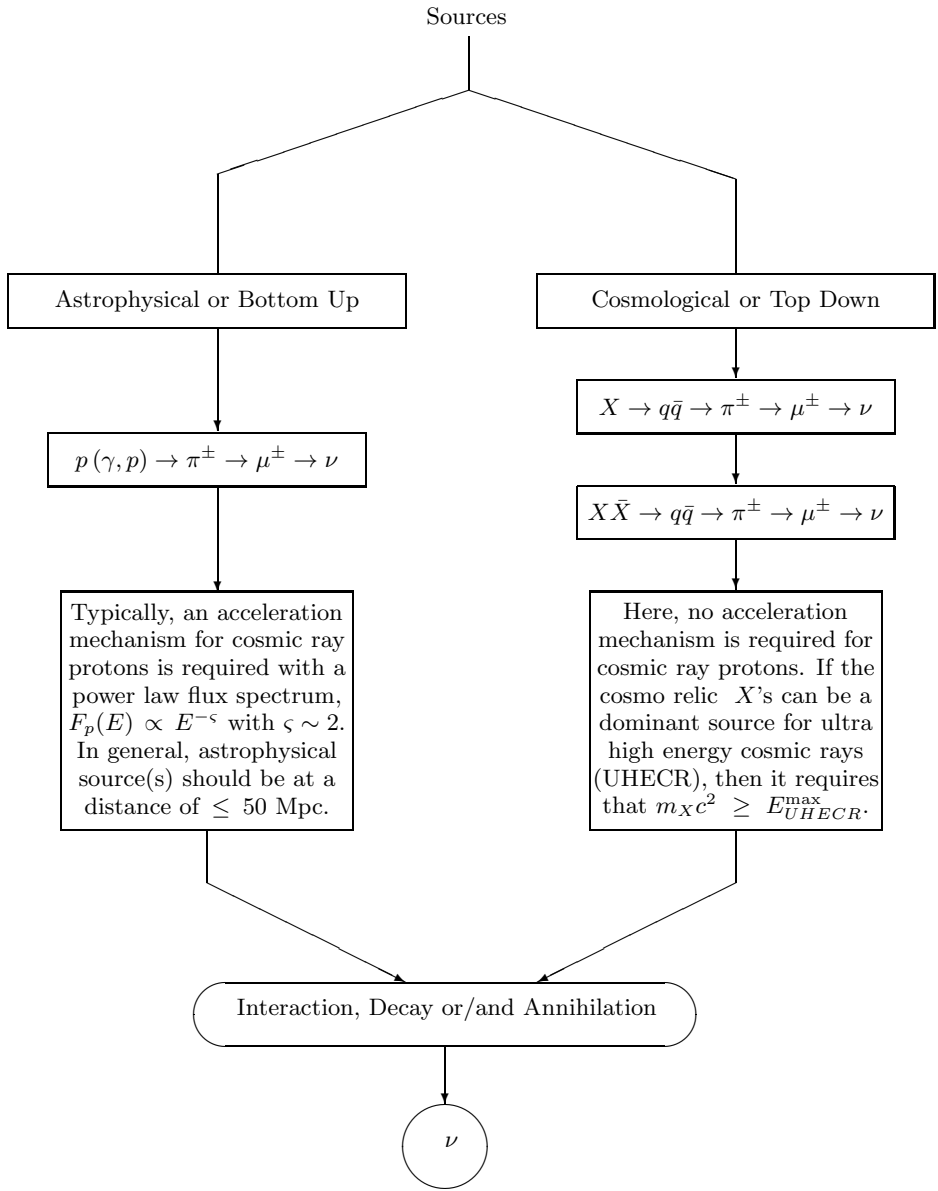
\begin{figure}
\label{figureone}
\begin{picture}(540,540)
\put(55,190){\framebox(110,90){\parbox{1.5in}{Typically, an acceleration mechanism 
 for cosmic ray protons is required with a power law flux spectrum, 
 $\, F_{p}(E)\, \propto \, E^{-\varsigma }$ 
 with $\varsigma  \sim 2$. 
 In general, astrophysical source(s) 
 should be at a distance of $\, \leq \, $ 50 Mpc.}}}
\put(110,190){\line(0,-1){20}}
\put(110,170){\vector(2,-1){75}}
\put(310,170){\vector(-2,-1){75}} 
\put(310,340){\vector(0,-1){19}}
\put(310,300){\vector(0,-1){19}}
\put(310,380){\vector(0,-1){19}}
\put(200,120){\oval(190,24){\makebox[1.11in][r]{Interaction, Decay or/and Annihilation}}} 
\put(200,70){\circle{30}{$\nu $}}
\put(200,108){\vector(0,-2){22}}
\put(50,320){\framebox(120,20){$p\, (\gamma,p)\to \pi^{\pm} \to \mu^{\pm}\to \nu$}}
\put(35,380){\framebox(150,20){Astrophysical or Bottom Up}}
\put(114,380){\vector(0,-1){39}}
\put(209,470){\line(-3,-1){95}} 
\put(209,470){\line(3,-1){101}}
\put(209,490){\line(0,-1){20}}
\put(310,436){\line(0,-1){36}}
\put(114,438){\line(0,-1){38}}
\put(114,320){\vector(0,-1){40}}
\put(193,495){Sources}
\put(235,380){\framebox(150,20){Cosmological or Top Down}}
\put(255,190){\framebox(110,90){\parbox{1.5in}{Here, no acceleration mechanism is
 required for cosmic ray protons.  
 If the cosmo relic $\, \, X$'s can be a dominant source for ultra high energy cosmic rays (UHECR), 
 then it requires that $m_{X}c^{2}\, \,  \geq \, \, E^{\rm max}_{UHECR}$.}}}
\put(250,340){\framebox(120,20){$X\to q\bar{q} \to \pi^{\pm}\to \mu^{\pm} \to \nu$}}
\put(310,190){\line(0,-1){20}}
\put(250,300){\dashbox{.40}(120,20){$X\bar{X}\to q\bar{q}
 \to \pi^{\pm}\to \mu^{\pm} \to \nu$}}
\end{picture}
\caption{A simple classification flow chart for presently envisaged main sources of
         high energy neutrinos. Only non tau neutrino production is illustrated.}
\end{figure}
    the observed high energy photon flux has purely
    electromagnetic origin. In the latter case, there will still
    be neutrino flux but at a rather suppressed level  
 (such as in $\gamma \gamma $ interactions) as compared
    to the former case. The latter possibility is recently
    discussed in some detail in \cite{Athar:2002ib}.

    The main interactions responsible for the production of these
    high energy neutrinos include the $p\gamma $ and $pp$  
interactions (see Table \ref{tableone} for some characteristics). For the behavior of these 
 cross sections as a function of center-of-mass energy 
$\sqrt{s}$ in the range of interest, see \cite{Hagiwara:fs}. There is formation of $\Delta $ 
 resonance in $p\gamma $ interactions,
 at  $\sqrt{s}\sim m_{\Delta}\sim 1.2$ GeV, where $s\simeq m^{2}_{p}+4E_{p}E_{\gamma}$, that mainly 
decay into electron and muon neutrinos. 
 Two behaviors of the $p\gamma $ cross section, near $\sqrt{s}\sim
1.2$  GeV  make it an important channel for high energy neutrino
production, the relatively large width of the $\Delta $ resonance,
 $\Gamma_{\Delta}/m_{\Delta} \sim 10^{-2}$, 
and the almost constant behavior of the cross section for $\sqrt{s}> m_{\Delta}+\Gamma_{\Delta}$.
Under the assumption of all other similar 
conditions, it is the interaction cross section that determines the absolute 
level of high energy neutrino production.

 For illustrative purpose, Fig. 1 displays a simple classification 
flow chart for presently envisaged sources of high energy neutrinos. 
It includes the possibility of high energy neutrino production 
from cosmic relics, referred to as $X$ \cite{Sigl:2001th}. Briefly, these relics are 
 considered to 
be formed in the early epochs of the universe such as during inflation epoch. 
The large amount of energy trapped in these relics may be 
released in the form of grand unification scale gauge bosons
 which in turn decay/annihilate into standard model particles 
 including neutrinos. 
 The direct channel $X\to \nu \bar{\nu}$ is also possible \cite{Berezinsky:2002hq}. 
 These relics need not be far away from us. In fact, 
some of the models suggest that they may be a part of our galactic dark matter halo 
 implying at a distance of $\leq $ 10 kpc. 
 If these $X$'s can be the dominant sources of
observed ultra high energy cosmic rays then this in turn severely 
constrain their number density $n_{X}$, life time $\tau_{X}$, mass $m_{X}$, and 
thus determine the resulting high energy neutrino flux spectrum shape and
absolute level. This possibility is referred to as 
the cosmological (or top down) scenario for expected high energy neutrino production.
 The distance restrictions do not necessarily always apply for this kind of
sources \cite{Weiler:1997sh}.
Currently, the ultra high energy cosmic rays with energy $E^{\rm max}_{UHECR}$ 
 up to $\sim 3\cdot 10^{11}$ GeV are observed \cite{Nagano:ve}. The 
cosmological sources however require physics beyond standard model to 
 work \cite{Berezinsky:2003iv}.

Depending on the details of the astrophysical or cosmological
model for high energy neutrino production scenario, either the
observed photon flux or proton flux or both are used to determine the
absolute level of the expected neutrinos flux.
From the cosmos, presently high energy photons and ultra high energy cosmic
rays (considered to be protons here) are observed in the relevant context. 
 Their observed level of flux  determines the absolute 
 flux level of neutrinos as high energy neutrinos are secondary in nature in the
sense that they are not matter particles and are not a significant
 fraction of the matter density associated with a specific known 
 astrophysical or/and cosmological  source. On
the other hand, neutrinos are stable and neutral and therefore for
this precise reason will carry useful information about the
source.

Supposing protons can escape the {\tt extra galactic} astrophysical sources and  
can be a dominant fraction of the observed ultra high energy cosmic ray
flux, the resulting high energy (muon) neutrino flux mainly in $p\gamma $ and $pp$ interactions 
either arising from inside the source or during propagation has to be less than
 this. It can be typically $\leq 10^{-8}\, {\rm GeV}({\rm cm}^{2}\cdot {\rm s \cdot sr})^{-1}$ 
  for
 $10^{5}< E/{\rm GeV} < 10^{12}$ \cite{Waxman:1998yy}. This bound further
 tightens by a factor of 1/2 once the neutrino flavor oscillation effects  are taken 
into account (see later).

	Consider now briefly the $p\gamma \, \to  \Delta \to p\pi\, (N=p)$ interactions 
	 occurring during 
 the propagation of ultra high energy cosmic rays either inside an 
 astrophysical source or between the source and the earth 
in the presence of  a dense photon background. This is to serve as an illustrative example
 for having an order of magnitude idea of the expected $E$. 
 The threshold energy for protons interacting with photons at an angle $\phi $ to form $\Delta $ resonance, is
\begin{equation}
    E^{\rm th}_{p}=
    \frac{(m_{p}+m_{\pi})^{2}-m_{p}^{2}}{2E_{\gamma}(1-\cos\phi)},
\end{equation}
which in case of head on interactions further simplifies to
\begin{equation}
\label{GZK}
    E^{\rm th}_{p}\simeq \frac{m_{p}m_{\pi}}{2E_{\gamma}}.
\end{equation}
 For $E_{p}<E^{\rm th}_{p}$, the interaction $p\gamma \to p e^{+}e^{-}$
 dominates the energy loss for protons. If $E_{\gamma}=E^{\rm CMB}_{\gamma}\sim 2.7$ K then 
$E^{\rm th}_{p}\sim  10^{11}$ GeV. 
 The $p\gamma $ interaction length can be defined
as
\begin{equation}
\lambda \sim 1/n_{\gamma}\sigma_{p\gamma \to p\pi}.
\end{equation}
For instance, if 
 $n_{\gamma}=n^{\rm CMB}_{\gamma} \sim 410$ cm$^{-3}$ for $E_{\gamma}=E^{\rm CMB}_{\gamma}$ 
 then $\lambda < 6$ Mpc, where $\sigma_{p\gamma \to p\pi}$ is given in Table I.

The propagation of ultra high energy proton flux, $F_{p}$ can be studied in 
 the presence of photon background in distance $r$, by
solving the following equation
\begin{equation}
    \frac{{\rm d}F_{p}(E,r)}{{\rm d}r}=-\frac{1}{\lambda(E)}F_{p}(E,r).
\end{equation}
The negative sign indicates the decrease in the ultra high energy proton
flux because of interaction described by $\lambda $. 
 This results in an exponential cut off in $p$ flux spectrum for $r \geq 50$ Mpc. In case of 
 ubiquitous cmb photon background, it is
 commonly referred to as Greisen Zatsepin Kuzmin (GZK) cut off \cite{Greisen:1966jv}. 
It occurs at $E^{\rm th}_{p}\sim 10^{11}$ GeV, according to Eq. (\ref{GZK}). 
The 
resulting GZK (muon) neutrino flux spectrum peaks at $\sim 10^{9}$ GeV by sharing 
roughly (1/4)$\cdot $(1/5) of the $E^{\rm th}_{p}$ \cite{Engel:2001hd}. 

The matter density in interstellar medium as well as in several of the astrophysical 
sites such as the galactic plane, the AGNs and the GRBs, is rather small 
 (relative to that in Atmosphere of 
earth). Therefore, a rather simple formula can be used to estimate the integral high 
energy neutrino flux spectrum in $p\gamma $ and/or $pp$, 
 in units of $({\rm cm}^{2}\cdot {\rm s \cdot sr})^{-1}$,  in a 
specific individual astrophysical site
\begin{equation}
\label{astroflux}
 F^{0}_{\nu}(E)=\int^{E^{\rm max}}_{E}
  {\mbox d}E\,  F_{p}\, (E)\,  g(E)
 \frac{\mbox{d}n_{p(\gamma ,p) \to \nu Y}}   {\mbox{d}E}.
\end{equation}
Here $F_{p}(E)$ parameterizes the high energy proton
flux. The function $g(E)\equiv r/\lambda (E) 
$ gives the number of $p (\gamma, p)$ interactions within the
 distance $r$. The ${\rm d}n/{\rm d}E\equiv \sigma^{-1}{\rm
d}\sigma/{\rm d}E$ is the neutrino energy distribution in above interactions.
The implicit assumption here is that the unstable hadrons and leptons produced 
in above interactions decay before they interact owing to the fact that the matter
density in the distance $r$ is assumed to be rather small. Also, the effects of possible 
 red shift
evolution and magnetic field  of the astrophysical sources are neglected for simplicity.

There is yet another possible class of astrophysical sources of high energy neutrinos
that are essentially neither constrained by observed high energy photon nor by ultra high
energy cosmic ray flux. It is so because in this class of sources, the
matter density is considered to be too large so that neither of the above
leave the source. These sources are therefore commonly referred to as hidden sources or 
 neutrinos only sources.
 The high energy neutrino production occurs in same $pp$ (or $p\gamma $) interactions here 
  also.  
 These can only be constrained by the high energy neutrino flux
(non) observations \cite{Berezinsky:2000bq}.

The above discussion is restricted to non tau neutrino production only. 
 In the $\pi^{\pm}\to \mu^{\pm} \to \nu$ 
decay situation, 
 the relative ratio of resulting electron and muon neutrino flux is 1 : 2 respectively. 
 The astrophysical tau neutrino flux is produced in decays of $D^{\pm}_{S}$. For 
 $\sqrt{s}\sim m_{\Delta} $, it is known that 
 $ \sigma[p(\gamma ,p)\to D^{\pm}_{S}Y]/ \sigma[p(\gamma ,p)\to \pi^{\pm}Y] 
 \leq {\cal O}(10^{-3}-10^{-4})$. 
 The high energy tau neutrino flux is thus rather suppressed at the production 
sites and can therefore be taken as approximately zero, 
 resulting in 1 : 2 : 0 \cite{Athar:1998ux}.
 For a recent review on astrophysical tau neutrinos, see \cite{Athar:2002rr},
whereas for cosmological tau neutrinos, see, for instance \cite{Wichoski:1998kh}.

\subsection{Oscillations during propagation: Effects of neutrino mixing}

In
view of recent growing evidence of neutrino flavor oscillations,
I shall here elaborate the {\tt relative} changes expected in the high energy neutrino
flux because of  neutrino flavor oscillations.

There are at least two aspects of neutrino propagation effects
that need somewhat careful considerations in study of neutrino 
 mixing effects for high energy neutrinos. These are: the neutrino 
interactions with the background particles inside the (astrophysical) source of
neutrinos as well as between the source and the earth.  
The present knowledge of matter density, $\rho $
inside the known sources as well as between these sources and the
 earth imply that it is rather quite small (as compared to that in Sun). 
 As a  result, the level crossing (or resonance) 
condition, namely $G_{F}\rho/m_{N}\sim \Delta m^{2}/2E$, for  matter enhanced neutrino flavor 
 oscillations is 
 not satisfied. 
 Level crossing is a necessary condition for occurrence of matter 
 enhanced neutrino flavor oscillations.  Therefore, there are
essentially no matter effects on pure vacuum flavor oscillations.
Note that this is in contrast to the situation in Sun and Super novae.  
 Furthermore, the neutrino nucleon and neutrino electron inelastic 
 interaction effects are also small enough to influence the mixed neutrino propagation even
at ultra high energy in a significantly observable manner. 
 This is also because of rather small matter density.   
 Therefore, I elaborate here only effects of neutrino flavor 
mixing in vacuum (with no matter interactions)\footnote{If 
i) $0.1\leq \sin^{2}2\theta \leq 0.95 $, ii) $E\geq 10^{19}$ GeV 
 (essentially irrespective of $\Delta m^{2}$ values), 
iii) the red shift $z\geq 3$ at production, and iv) $\xi \geq 1$, where 
 $\xi\equiv (n_{\nu}-n_{\bar{\nu}})/n_{\gamma}$, then a deviation from pure 
vacuum flavor oscillations can be of the order of few percent, when high energy 
 neutrinos
scatter over the very low energy relic neutrinos during their propagation to us 
 in the interstellar medium \cite{Lunardini:2000fy}.}.

Note from the previous subsection that the high 
 energy neutrinos
are produced in the following relative ratios
\begin{equation}
\label{initial}
F^{0}_{\nu_{e}}: F^{0}_{\nu_{\mu}}: F^{0}_{\nu_{\tau}}=1: 2: 0.
\end{equation}
It is assumed here that the high energy neutrinos and anti neutrinos originate
in equal proportion from a source and are counted in the symbol $\nu $ together. 
 Also as the absolute level of 
high energy neutrino flux is presently unknown, I therefore elaborate
the neutrino mixing effects on relative ratios only, in the
context of three flavors. 
 Four flavor mixing effects are considered in 
 \cite{Athar:2000yw,Athar:2000tg}.

To obtain a general expression for flavor oscillation formula, I
start with the connection $U$ between the flavor $|\nu_{\alpha}\rangle $
 and mass $|\nu_{i}\rangle$ eigen 
states of neutrinos, namely
\begin{equation}
|\nu_{\alpha}\rangle =\sum^{3}_{i = 1}U_{\alpha i}|\nu_{i}\rangle,
\end{equation}
where $\alpha = e, \mu $ or $\tau $. 
 In the context of three neutrinos, $U$ is
called Maki Nakagawa Sakita (MNS) mixing matrix \cite{Maki:mu}.
  It can be
obtained by performing the following operations to coincide with the 
one given in \cite{Hagiwara:fs}: 
\begin{equation}
 U\equiv R_{23}(\theta_{23})\cdot {\rm diag}(e^{-i\delta_{13}/2},
 1,\, e^{i\delta_{13}/2})\cdot R_{13}(\theta_{13})\cdot
 {\rm diag}(e^{i\delta_{13}/2},1,\, e^{-i\delta_{13}/2})
 \cdot R_{12}(\theta_{12}),
\end{equation}
where $\theta$'s are neutrino mixing angles and $\delta_{13}$ is CP violation phase.
Explicitly, it reads
\begin{equation}
 U=\left( \begin{array}{ccc}
          c_{12}c_{13} & s_{12}c_{13} & s_{13}e^{-i\delta_{13}}\\
          -s_{12}c_{23}-c_{12}s_{23}s_{13}e^{i \delta_{13}} & 
          c_{12}c_{23}-s_{12}s_{23}s_{13}e^{i\delta_{13}} & 
          s_{23}c_{13}\\
          s_{12}s_{23}-c_{12}c_{23}s_{13}e^{i\delta_{13}} & 
          -c_{12}s_{23}-s_{12}c_{23}s_{13}e^{i\delta_{13}} & 
          c_{23}c_{13}
          \end{array}
   \right).
\end{equation}
Here $c_{ij}=\cos \theta_{ij}$ and $s_{ij}=\sin \theta_{ij}$ (with $j=1,2,3$) 
 and that $UU^{\dagger}=1$.
Using it, one obtains the following well known formula for flavor
 oscillation probability from $\alpha$ to $\beta \, (\beta =e, \mu \, {\rm or}\, \tau) $ 
 for a neutrino source at a fixed distance $L$: 
\begin{equation}
\label{complete}
P(\nu_{\alpha} \to \nu_{\beta}; L) \equiv P_{\alpha \beta}=  
 \delta_{\alpha \beta}-\sum_{j\neq k}
 U^{*}_{\alpha j}U_{\beta j}U_{\alpha k}
 U^{*}_{\beta k} (1-e^{-i\Delta m^{2}_{jk}L/2E}). 
\end{equation}
In the far distance approximation, namely, in the limit $L\to \infty$, one obtains
\begin{eqnarray}
\label{osc-prob}
 P(\nu_{\alpha} \to \nu_{\beta}; L\to \infty)& \simeq & 
 \delta_{\alpha \beta} -\sum_{j \neq k} 
 U^{*}_{\alpha j}U_{\beta j}U_{\alpha k}
 U^{*}_{\beta k},\nonumber \\ 
  & \simeq  & 
 \sum^{3}_{j =1} |U_{\alpha j}|^{2}
 |U_{\alpha j}|^{2}.
\end{eqnarray}
Because of the assumed averaging over the rapidly oscillating phase
 ($l_{\rm osc}\ll L, {\rm where} \, \, l_{\rm osc}\equiv 2E/\Delta m^{2}_{jk}$), 
  the last two expressions
 are {\tt independent} of $E$  and $\Delta m^{2}$. Under this assumption, the oscillation 
probability can be written as a symmetric matrix $P$ such that
\begin{equation}
 P=\left( \begin{array}{ccc}
          P_{ee} & P_{e\mu } & P_{e\tau}\\
          P_{e\mu } & P_{\mu \mu } & P_{\mu \tau}\\
	    P_{e\tau } & P_{\mu \tau} & P_{\tau \tau} 
          \end{array}
   \right).
\end{equation}
For vanishing $\delta_{13} $ and $\theta_{13}$, 
 using Eq. (\ref{osc-prob}), it is straight forward to obtain
\begin{equation}
 P_{e\mu } \simeq 2c^{2}_{12}s^{2}_{12}c^{2}_{23}, \, \, \, 
 P_{e\tau } \simeq 2c^{2}_{12}s^{2}_{12}s^{2}_{23}, \, \, \, {\rm and}\, \, \, 
P_{\mu \tau} \simeq c^{2}_{23}s^{2}_{23}(1+c^{4}_{12}+s^{4}_{12}). \, \, \, 
\end{equation}
Using above Eq., a  simple form for $P$ matrix can be obtained  in case of
 bi maximal mixing as \cite{Athar:1999at}:
\begin{equation}
 P=\left( \begin{array}{ccc}
          5/8  & 3/16  & 3/16\\
          3/16 & 13/32 & 13/32\\
          3/16 & 13/32 & 13/32
          \end{array}
   \right),
\end{equation}
as presently in the context of three (active) neutrino flavors,
 the Solar electron neutrino deficit can be explained with 
 $(\Delta m^{2}, \sin^{2}2\theta)$ as $(10^{-5}\, {\rm eV}^{2}, \sim 1)$ via
 $\nu_{e}\to \nu_{\mu} \, \, {\rm or}\, \,  \nu_{\tau}$ oscillations. 
 The Atmospheric muon neutrino 
deficit can be explained with $(10^{-3}\, {\rm eV}^{2}, \sim 1)$ via
 $\nu_{\mu}\to  \nu_{\tau}$ oscillations. 
The above $P$ matrix satisfies the following unitarity
conditions:
\begin{equation}
 1-P_{ee}=P_{e\mu}+P_{e\tau}, \, \, \, 
 1-P_{\mu \mu}=P_{e\mu}+P_{\mu \tau}, \, \, \, {\rm and}\, \, \, 
 1-P_{\tau \tau}=P_{e\tau}+P_{\mu \tau}.
\end{equation}
Namely, the disappearance of a certain neutrino flavor is equal to 
the appearance of this flavor into other (active) neutrino flavors. 
High energy neutrino flux arriving at the earth can then be estimated 
using 
\begin{equation}
\label{atearth}
 F_{\nu_{\alpha}}=\sum_{\beta}P_{\alpha \beta}F^{0}_{\nu_{\beta}},
\end{equation}
where $P_{\alpha \beta}$ is given by Eq. (\ref{osc-prob}). Note that in case
of initial relative flux ratios as 1 : 2 : 0 [see Eq. (\ref{initial})], one always get 
\begin{equation}
 F_{\nu_{e}}: F_{\nu_{\mu}}: F_{\nu_{\tau}}=1: 1: 1,
\end{equation}
under the assumption of
averaging {\tt irrespective} of any specific flavor oscillation 
 solution for Solar neutrino problem \cite{Athar:2000je}.
 A considerable enhancement in $F_{\nu_{\tau}}$ 
 relative to $F^{0}_{\nu_{\tau}}$ because of neutrino oscillations is evident.   
  A some what detailed numerical study that takes into account 
 the effects of non vanishing $\delta_{13}$ and $\theta_{13}$ indicates that the 
 deviation $\epsilon $,  from these 
final relative ratios is not more than few percent 
 (namely, $|\epsilon | \leq 0.1$ in $1\pm |\epsilon |$) \cite{Athar:2000yw}.   
  There could, in principle, be several intrinsic neutrino properties that may
lead to {\tt deviations} from 1 : 1 : 1 final relative ratios other than $|\epsilon |$ as well as 
 an {\tt energy dependence}, such as neutrino spin flavor conversions \cite{Athar:1999gw}. 
 Some astrophysical/cosmological reasons at the source can also possibly 
 contribute to $\epsilon $. 

Measuring the three flavor ratios (and in particular deviations from 1 : 1 : 1) may 
entail several important consequences such as $\nu_{\tau}$ flavor appearance (which
has not yet achieved in terrestrial experiments because of flavor oscillations), 
the insight into the production mechanism (whether through $\pi^{\pm}$ or not) and
about its astrophysical or cosmological origin \cite{Gounaris:2002ab}.

In the  above simplified discussion, the expression for $P$ neither depends on $\Delta
m^{2}$ nor on $E $. However, in some situations, this need not be the case. 
 In that case, one need to use complete expression for
$P$ given by Eq. (\ref{complete}) and possibly have to average over the red shift 
 distribution of astrophysical sources, $f(z)$. 
 This gives the effect of
evolution of the sources with respect to $z$. The $P$ can then be calculated
using Eq. (\ref{complete}) with $E \to (1+z)E$ in 
following formula
\begin{equation}
 P_{\alpha \beta}(E)=\frac{\int^{z_{max}}_{0}P_{\alpha \beta}(E,z) f(z) {\rm d}z}
 {\int^{z_{max}}_{0} f(z) {\rm d}z}.
\end{equation}
The $f(z)$ can be found in \cite{Lunardini:2000fy}.	
\subsection{Prospects for possible future observations}
I shall here describe
the basic essential factors such as the neutrino nucleon/electron 
 interaction cross section and the range of the associated charged leptons 
 typically in the energy range $10^{3\div 4} < E/{\rm GeV} < 10^{7}$ 
 that determine the (limited) near future
prospects for observations of high energy neutrinos. 
 The current and near future status of the dedicated high energy neutrino detectors
is given in \cite{Spiering:2003xm}.

Briefly, 
 the present detectors based on Cherenkov radiation measurement, in ice or water are 
 the Antarctic Muon and Neutrino Detector Array (AMANDA) and 
 its proposed extension, the Ice Cube, the lake Baikal detector and the Astronomy with 
 a Neutrino Telescope and Abyss environmental RESearch (ANTARES) detector array. 
The hybrid detectors based on particle and radiation measurement such as Pierre 
Auger Observatory may also detect high energy neutrinos \cite{Capelle:1998zz}. 
 Detectors based on alternative
detection techniques such as radio wave detection are also in operation, such as
Radio Ice Cherenkov Experiment (RICE). This detector is based 
 on Askaryan effect \cite{askaryan}. This effect is briefly defined as follows: 
 In an electromagentic shower generated in deep inelastic neutrino nucleon 
 interaction, the electrons and photons in the shower generate an excess 
 of $\sim 10-20 \, \%$ electrons in the shower because of the electron and 
 photon interactions with the medium in which the shower develops. 
 This in turn generate  coherent radio wave pulse (in addition to other type 
 of radiation), if the wavelength of this radio emission is greater 
 than the size of the shower.
	The search for alternative high energy neutrino detection 
 medium other than air, water and/or ice, such as rock salt  
 has also been attempted for radio wave emission \cite{Chiba:2001gx}. 
 It might also be possible to detect the
 acoustic pulses generated by deep inelastic neutrino nucleon interactions near or inside the 
 detector. An attempt in this direction is through 
 Sea Acoustic Detection of Cosmic Objects (SADCO) detector array.
  Other modern proposals include space based detectors such as Orbiting Wide Angle Light collector 
 (OWL/Air Watch)
 and Extreme Universe Space Observatory (EUSO).

 Among all these, somewhat stringent upper bounds on (diffuse) 
 high energy neutrino flux are reported by 
 AMANDA and Baikal 
detectors. From AMANDA (B10), now it is typically 
 $\leq 8.4\times 10^{-7}\, {\rm GeV} ({\rm cm}^{2}\cdot {\rm s}\cdot {\rm sr})^{-1}$ in the energy
range $6\cdot 10^{3}< E/{\rm GeV} < 10^{6}$ \cite{Ahrens:2003ee}.  
 The effective area for AMANDA detector 
 is $\sim 10^{-2}$ 
 km$^{2}$ for a $10^{4}$ GeV muon neutrino. This upper bound is based on 
 non observations of upward going high energy muon neutrinos (with $\sim E^{-2}$ energy spectrum 
 index) after subtracting the
 relevant Atmospheric muon neutrino background. The next generation of high energy neutrino 
 detectors are considered to have an effective area of $\sim $ 1 km$^{2}$.

 The high energy neutrino observation can be achieved in the
 following two main interactions: the deep inelastic neutrino nucleon and neutrino
 electron interactions. The deep inelastic neutrino nucleon interaction can 
 proceed via $Z$ or $W^{\pm}$ exchange. The former is 
 called neutral current (NC) interaction, whereas the later is called charged current (CC)
 interaction. The CC interactions $(\nu_{\alpha}N\to \alpha Y)$ are most relevant for 
  prospective high 
 energy neutrino observations. The showers, the charged particles and the 
 associated radiation emission such as Cherenkov radiation from these interactions 
 are the measurable  quantities. 
 The CC deep inelastic neutrino
 nucleon cross section $\sigma^{CC}_{\nu_{\alpha}N}(E)$ over nucleons  with mass $m_{N}$,  
 can be written as a function of  the incoming neutrino energy $E$ as:
\begin{eqnarray}
\label{DIS}
 \sigma^{\rm CC}_{\nu_{\alpha}N}(E) & = & \frac{2G^{2}_{F}m_{N}E}{\pi}\int {\rm d}x \int {\rm d}y
 \left (\frac{M^{2}_{W}}{Q^{2}+M^{2}_{W}}\right)^{2} \cdot x \cdot
 \nonumber \\
 & &
 \left \{ \left(1-\frac{m^{2}_{N}x^{2}y^{2}}{Q^{2}}\right)
 [f_{d}(x, Q^{2})+f_{s}(x, Q^{2})+f_{b}(x, Q^{2})]+ \right. \nonumber \\
 & & \left. \left( (1-y)^{2}-\frac{m^{2}_{N}x^{2}y^{2}}{Q^{2}}\right)
 [f_{\bar{u}}(x, Q^{2})+f_{\bar{c}}(x, Q^{2})+f_{\bar{t}}(x, Q^{2})]\right\}.
\end{eqnarray}
The integration limit for $x$ and $y$ can be taken between 0 and 1. 
This expression can be straight forwardly obtained using $s=2m_{N}E$ 
 in \cite{Hagiwara:fs}. Here $f_{q}(x, Q^{2})$
are the parton distribution functions. The $f_{q}(x, Q^{2})$'s may
be evaluated at $Q^{2}=m^{2}_{Z}$ for small $x$ ($x\leq 10^{-4\div -5}$).
The $-Q^{2}$ is the invariant momentum transfer between the 
 incoming neutrino and outgoing charged lepton. 
 In above Eq., $y\equiv (E-E^{\prime})/E$ is the 
 inelasticity in the neutrino nucleon interactions. 
 It gives the fraction of $E$ 
lost in a single neutrino nucleon interaction in the lab frame. 
 The $x \equiv Q^{2}/2m_{N}(E-E^{\prime})$ is the fraction of 
 the nucleon's momentum carried by the struck quark.
 The charged lepton mass is ignored here in comparison with $m_{N}$. 
The $\sigma^{\rm CC}_{\bar{\nu}_{\alpha}N}(E)$ for anti neutrinos can be obtained 
using Eq. (\ref{DIS}) with appropriate changes.

\begin{figure}
\includegraphics[width=2in]{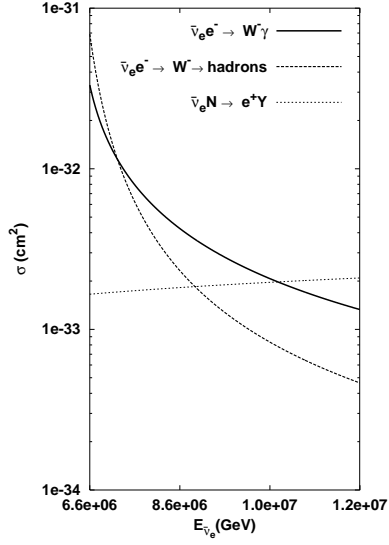}
\caption{Examples of high energy $\bar{\nu}_{e}$ interaction cross section 
 over two different target particles as a function of $E_{\bar{\nu}_{e}}$ (GeV).
 The minimum value of $E_{\bar{\nu}_{e}}$ corresponds to $(m_{W}+\Gamma_{W})^{2}/2m_{e}$.}
\label{fig}
\end{figure}

 The neutrino electron interaction cross section 
 on the other hand has a resonant character for $s=2 m_{e}E_{\bar{\nu}_{e}}$:
\begin{equation}
\label{sigma}
 \sigma(\bar{\nu}_{e}e \to W^{-} \to {\rm hadrons})
 =\frac{\Gamma_{W}({\rm hadrons})}{\Gamma_{W}({e^{+}\nu})}\cdot \frac{G^{2}_{F}s}{3\pi}
 \cdot \left[\frac{m^{4}_{W}}
 {(s-m_{W}^{2})^{2}+\Gamma^{2}_{W} m^{2}_{W}}\right],
\end{equation}
where $\Gamma_{W}$'s can be found in \cite{Hagiwara:fs}. The
above resonant interaction select the anti electron neutrino flavor as well as the
energy, namely $E_{\bar{\nu}_{e}}=m^{2}_{W}/(2m_{e})\sim 6.3\cdot 10^{6}$ GeV. 
 This interaction may, in principle, be used to {\tt calibrate} the 
 high energy neutrino energy  provided it can possibly be discriminated from neutrino nucleon 
 interaction in a detector. The 
 $\sigma(\bar{\nu}_{e}e \to W^{-} \to {\rm hadrons})$
 has a slight enhancement because of hard photon emission in the 
final state for $\sqrt{s}\geq \Gamma_{W}$. It is given by \cite{Athar:2001bi}
\begin{equation}
\label{six}
 \sigma(\bar{\nu}_{e}e^{-}\rightarrow W^{-}\gamma)=
 \frac{\sqrt{2}\alpha G_{F}}{3u^{2}(u-1)} \left[3(u^{2}+1)\ln
 \left\{\frac{(u-1)m^{2}_{W}}{m^{2}_{e}}\right\}
       -(5u^{2}-4u+5)\right],
\end{equation}
where $u=s/m^{2}_{W}$. 
In Fig. \ref{fig}, the three cross sections are plotted for illustration. 
The $\sigma^{\rm CC}_{\bar{\nu}_{e}N}(E)$ is calculated using Eq. (\ref{DIS}) 
 with CTEQ(5M) parton 
 distribution functions generated by Coordinated Theoretical and Experimental Project 
 on QCD Phenomenology and Tests of the Standard Model \cite{Lai:1994bb}.

The high energy neutrino flux arrives at an earth based detector in three 
general directions in {\tt equal proportion}. 
The downward going neutrinos do not cross any significant earth cord before reaching
the (under ground) detector. The (quasi) horizontal and upward going neutrinos 
 cross the earth with increasing 
 cord length respectively before reaching the detector. 

\subsubsection{Downward going}
The event rate for downward going high energy neutrinos in CC deep 
inelastic interactions is given by \cite{Gandhi:1998ri}
\begin{equation}
    {\rm Rate}= A \int^{E^{\rm max}_{\nu_{\alpha}}}_{E^{\rm min}_{\alpha}} {\rm
    d}E_{\nu_{\alpha}} P_{\nu_{\alpha}\to \alpha} (E_{\nu_{\alpha}}, E^{\rm min}_{\alpha})
    F_{\nu_{\alpha}},
\end{equation}
here $A$ is the area of the high energy neutrino detector. 
 The $F_{\nu_{\alpha}}$ can be obtained using Eq. (\ref{atearth}). In the above
equation
\begin{equation}
\label{prob}
 P_{\nu_{\alpha }\to \alpha } (E_{\nu_{\alpha }}, E^{\rm min}_{\alpha })= N_{A}
  \int^{1-E^{\rm min}_{\alpha }/E_{\nu_{\alpha }}}_{0}
  {\rm d}y R_{\alpha }(E_{\nu_{\alpha }}, E^{\rm min}_{\alpha })
  \frac{{\rm d}\sigma^{\rm CC}_{\nu_{\alpha }N}(E_{\nu_{\alpha }}, y)}{{\rm
  d}y}.
\end{equation}
The ${\rm d}\sigma/{\rm d}y$ can be obtained using Eq. (\ref{DIS}).  
 The $N_{A}$ is the Avogadro's constant. Various $R$'s are given in
Table \ref{table2}. Note that for $\tau $ lepton,
 it is the decay length that is considered as its range
 with $E^{\rm min}_{\tau}\sim 2\cdot 10^{6}$ GeV and 
 $E^{\rm max}_{\nu_{\tau}}\sim 2\cdot 10^{7}$ GeV as
 the value of $D$ is chosen as 10$^{5}$ cm for illustration 
 here \cite{Athar:2000rx}.
Also note that $R_{e}\equiv R_{e}(E)$ only. 

\begin{table}[b]
\caption{The three charged lepton ranges discussed in the text.}
\label{table2}
\begin{tabular}{|c|c|c|}
\hline
\hline
Lepton Flavor & $R $(cmwe) & Ref. \\
\hline
 $e$& 40 $\left[(1-\langle y(E) \rangle) \frac{E}{6.4\cdot 10^{4}{\rm GeV}}\right]$ 
 & \cite{Gandhi:1998ri}\\
 $\mu$ & $\frac{1}{b}\ln \left( \frac{a+bE_{\mu}}{a+bE^{\rm min}_{\mu}}\right)$,
 $a = 2\cdot 10^{-3}\, \, \, $GeV/cmwe, b =  $3.9\cdot 10^{-6}$ /cmwe & 
 \cite{Gandhi:1998ri}\\
 $\tau$ & $D - \frac{E(1-y)\tau_{c}}{m_{\tau}c^{2}}$, D = $10^{5}{\rm cm}$ & 
 \cite{Athar:2000rx}\\
\hline
\hline
\end{tabular}
\end{table}

\begin{figure}[t]
\hglue -8.cm
\includegraphics[width=4in]{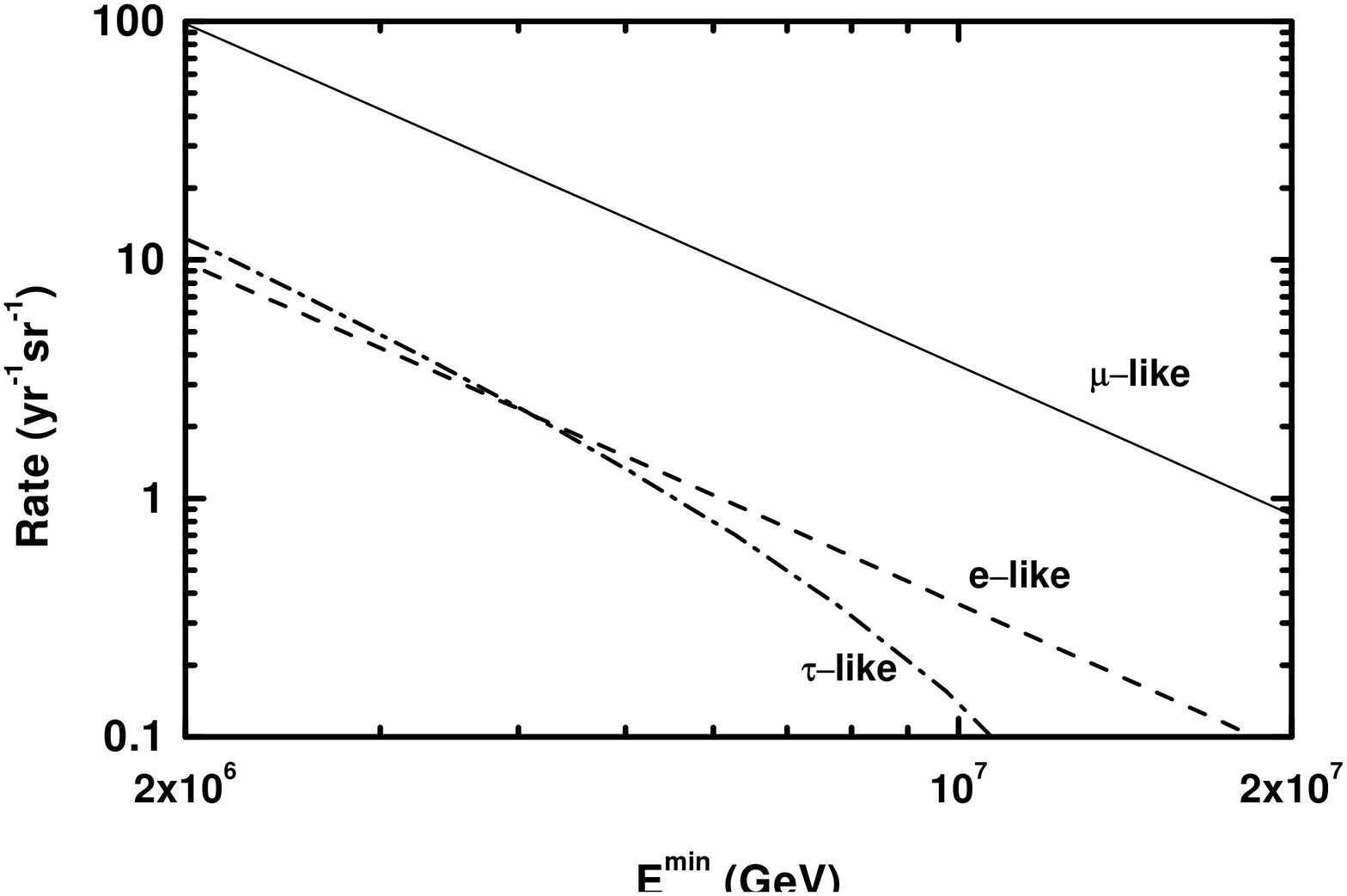}
\vglue -7.0cm 
\hglue 8.8cm \includegraphics[width=2.8in]{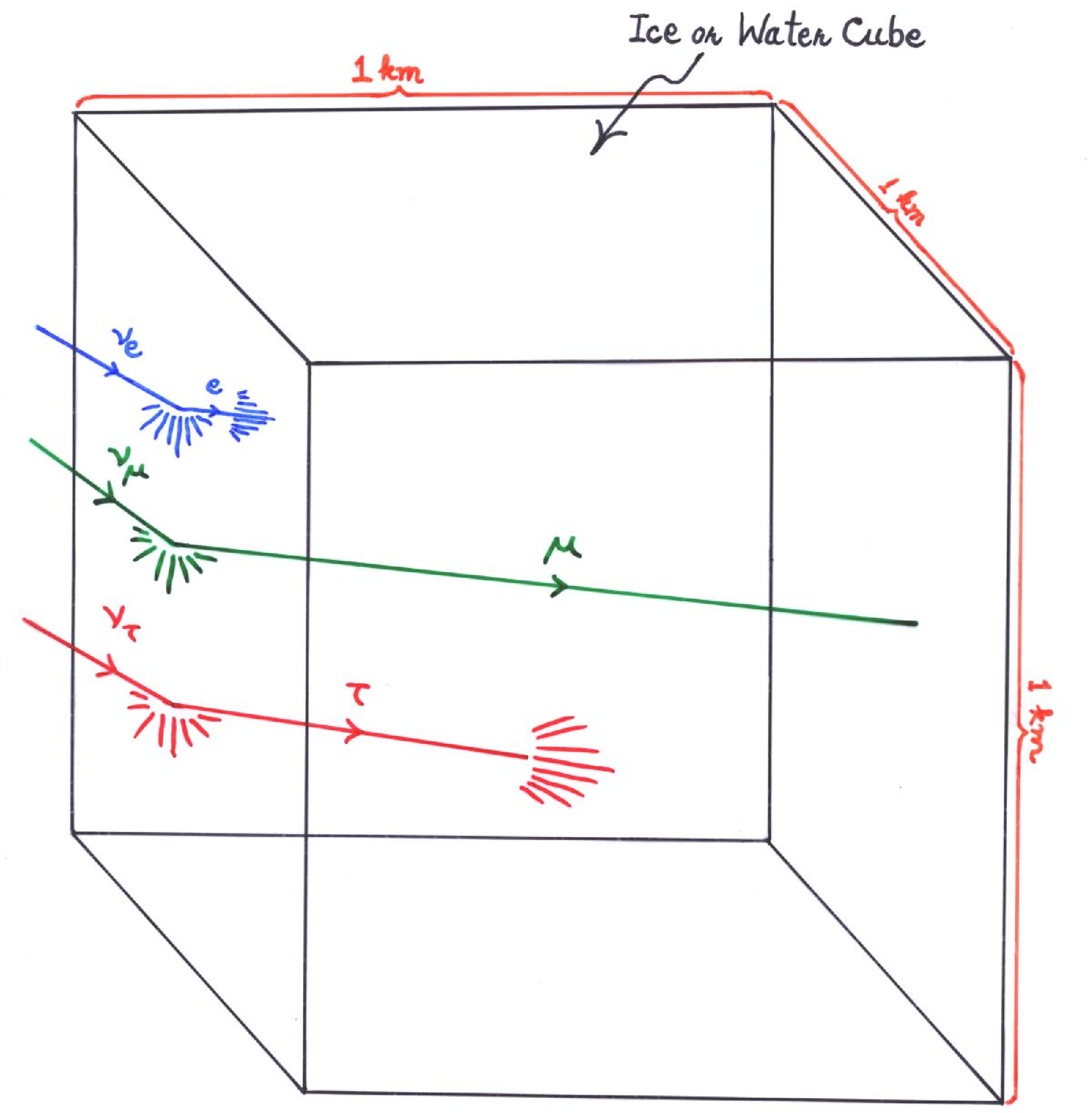}
\caption{Left panel: Expected downward going $e-$like,
 $\mu-$like and $\tau-$like event rate produced by AGN neutrinos as
 a function of minimum energy of the corresponding charged lepton
 in a large (proposed) km$^{3}$ volume ice or water neutrino detector. Three
 flavor neutrino mixing is assumed.
 Right panel: 
 Approximate representative event topologies for the three neutrino 
 flavors in a 
 km$^{3}$ volume water or ice neutrino detector for the order of 
 magnitude energy interval shown in left panel.}
\label{eventrate}
\end{figure}

Detailed estimates of the high energy neutrino  event rates are
done mainly numerically \cite{Giesel:2003hj}. 
 These estimates  are model
dependent. The event rates of  downward high energy 
neutrinos  typically vary between $\sim \, {\cal O}(10^{1})$
and
 $\sim \, {\cal O}(10^{2})$ in units of (yr  sr)$^{-1}$ for the proposed 
 km$^{3}$ volume ice or water neutrino detector.
The left panel of 
 Fig. \ref{eventrate} displays the three downward going 
event rates along with examples of approximate event topologies 
 for  AGN neutrinos  \cite{Szabo:qx}.
 In this AGN model, the $pp$  interactions inside the core of the AGN are  
 considered to play an important role. 
The $e-$ like event rate is obtained by re scaling the
$\mu-$ like event rate. The indicated order of magnitude 
energy interval is relevant for proposed km$^{3}$ volume high energy 
 neutrino detectors for possible neutrino flavor 
 identification.

The downward going  high energy neutrinos of different
flavors interact with the medium (free nucleons) of the detector,
deep inelastically mainly through CC 
interactions. The three flavors on the average give rise to {\tt
different} event topologies
 based on
these interactions and the behavior of the associated 
{\tt charged lepton}.
 For instance, for $10^{6}\leq E/{\rm GeV} \leq 10^{7}$, in proposed km$^{3}$ volume
 ice or water  
 neutrino detectors, typically the downward going high energy
 electron neutrinos produce a single shower, the muon neutrinos
produce muon like tracks passing through the detector (along with a single shower), 
 whereas the
 tau neutrinos produce two
 hadronic showers connected by tau (muon like) track and is such that the amplitude
  of the second shower is essentially a factor of two larger as compared to the first.
 Here, amplitude refers to maximum number of charged particles per unit
 length (see right panel of Fig. \ref{eventrate}) \cite{Learned:1994wg,Athar:2000rx}. 
\subsubsection{Upward going}
 For upward going high energy neutrinos, a shadow factor $S(E)$
is included in the integral given by Eq. (\ref{prob}). 
 The shadow factor $S(E)$ takes into account the effects 
 of absorption by earth \cite{Gandhi:1998ri}.
 The absorption of upward going high energy neutrinos by earth is
 {\tt neutrino flavor
dependent}. For $E \, \geq 10^{6}$ GeV, the upward  going tau
neutrinos
 may reach the
surface of the earth in a relatively small number by lowering their energy so that
 $E \, < 10^{6}$ GeV \cite{Halzen:1998be},
whereas the upward going electron and muon neutrinos are almost
completely
 absorbed by the earth. 
 For further details,
see \cite{Gandhi:1998ri}.  
\subsubsection{Quasi Horizontal}
It might become possible to observe the (air) showers produced 
by charged leptons produced in  
 neutrino nucleon interactions 
occurring near the surface of earth. Here a cord of earth
essentially equal to just one neutrino nucleon interaction length
is considered to be traversed by high energy neutrinos before reaching the detector.
This situation is in {\tt contrast} to upward going high energy 
 neutrinos which are basically completely absorbed 
 by the earth because of multiple interactions \cite{Fargion:2000iz}.

For a recent discussion on prospects for observations of 
near horizontal high energy (tau)  neutrinos, 
see, \cite{Iyer:1999wu}, whereas for 
ultra high energy neutrinos, see \cite{Feng:2001ue}. The 
emerging tau lepton spectra induced by quasi horizontal 
(or earth skimming) neutrinos, using Monte Carlo simulation 
 techniques are calculated in some detail in \cite{Bottai:2002nn}.
\section{Summary and Conclusions}
Detailed study of high energy neutrino fluxes from different astrophysical sites such as the
 galactic plane, as well as other more far away 
anticipated astrophysical and cosmological 
 sites will provide valuable information about the neutrino intrinsic
properties and the site itself.  Given the current status of absolute 
level of expected high energy neutrinos, the detectors with area $> (0.1-1)
\, \, {\rm km}^{2}$ seems to be needed to obtain the first evidence.

For more distant and energetic 
 sources, 
the prospective high energy neutrino observation will provide clues for a
solution of the long standing problem of origin of observed ultra high energy 
cosmic rays. The (non) observation of high energy neutrinos will also help to
better model the underlying physics of the far away astrophysical and 
cosmological  sites. In this context, present
motivations for their searches are reviewed, with a 
 description of their possible connection 
with ultra high energy cosmic rays. Some presently envisaged main high energy neutrino production 
mechanisms are summarized via a simple classification flow chart. Three
flavor neutrino oscillation description is reviewed and its implications
for the three relative ratios of high energy neutrinos are given. Furthermore,
 the essentials of prospective observations of high energy neutrinos are briefly 
described including the possible relevant observational signatures.     
 
\section*{Acknowledgement}  

This work is supported by Physics Division of NCTS.

\end{document}